\documentclass[submission,copyright,creativecommons]{eptcs}
\providecommand{\event}{FMAS'22} 

\usepackage{graphicx}
\usepackage{wrapfig}
\usepackage{iftex}

\ifpdf
  \usepackage{underscore}         
  \usepackage[T1]{fontenc}        
\else
  \usepackage{breakurl}           
\fi

\usepackage{ifthen,comment,amssymb, amsmath}
\usepackage{xcolor}
\usepackage{enumitem}

\newboolean{showcomments}
\setboolean{showcomments}{true}
\ifthenelse{\boolean{showcomments}}
 { \newcommand{\mynote}[3]{
      \fbox{\bfseries\sffamily\scriptsize#1}
        {\small$\blacktriangleright$\textsf{\textcolor{#3}{{\em #2}\bf }}$\blacktriangleleft$}}}
        { \newcommand{\mynote}[2]{}}

\usepackage{todonotes}

\title{Towards a Digital Highway Code using Formal Modelling and Verification of Timed Automata}

\author{Gleifer Vaz Alves
\institute{Federal University of Technology\\Parana, Brazil}
\email{gleifer@utfpr.edu.br}
\and
Maike Schwammberger\thanks{M.\ Schwammberger was supported by the German Research Council (DFG) in the PIRE Projects SD-SSCPS and ISCE-ACPS under grant no.\ FR 2715/4-1 and FR 2715/5-1.}
\institute{University of Oldenburg \\Oldenburg, Germany}
\email{schwammberger@informatik.uni-oldenburg.de}
}
\def\titlerunning{Formal Modelling and Verification of Traffic Rules}
\def\authorrunning{Gleifer Vaz Alves, Maike Schwammberger}
\begin{document}
\maketitle

\begin{abstract}
One of the challenges in designing safe, reliable and trustworthy Autonomous Vehicles (AVs) is to ensure that the AVs abide by traffic rules. For this, the AVs need to be able to understand and reason about traffic rules. In previous work, we introduce the spatial traffic logic USL-TR to allow for the unambiguous, machine-readable, formalisation of traffic rules. This is only the first step towards autonomous traffic agents that verifiably follow traffic rules. In this research preview, we focus on two further steps: a) retrieving behaviour diagrams directly from traffic rules and b) converting the behaviour diagrams into timed automata that are using formulae of USL-TR in guards and invariants. With this, we have a formal representation for traffic rules and can move towards the establishment of a Digital Highway Code. We briefly envision further steps which include adding environment and agent models to the timed automata to finally implement and verify these traffic rule models using a selection of formal verification tools.
\end{abstract}

\section{Motivation}\label{sec:introduction}
Our endeavor is to support the inclusion of traffic rules into the design process of Autonomous Vehicles (AVs). For this, we envision a Digital Highway Code, which is a machine-readable version of the natural language traffic rules as they exist for human drivers in various languages for various countries \cite{departmentForTransportUsing2017, stvo}. Using such a Digital Highway Code, AVs would be able to understand and reason about traffic rules. With this, explainability of AV behaviour would be supported, e.g. if and why a traffic rule was violated.

\begin{figure}[htbp]
\centering
    \includegraphics[width=0.8\linewidth,trim=1cm 0.0cm 0 0.0cm, clip]{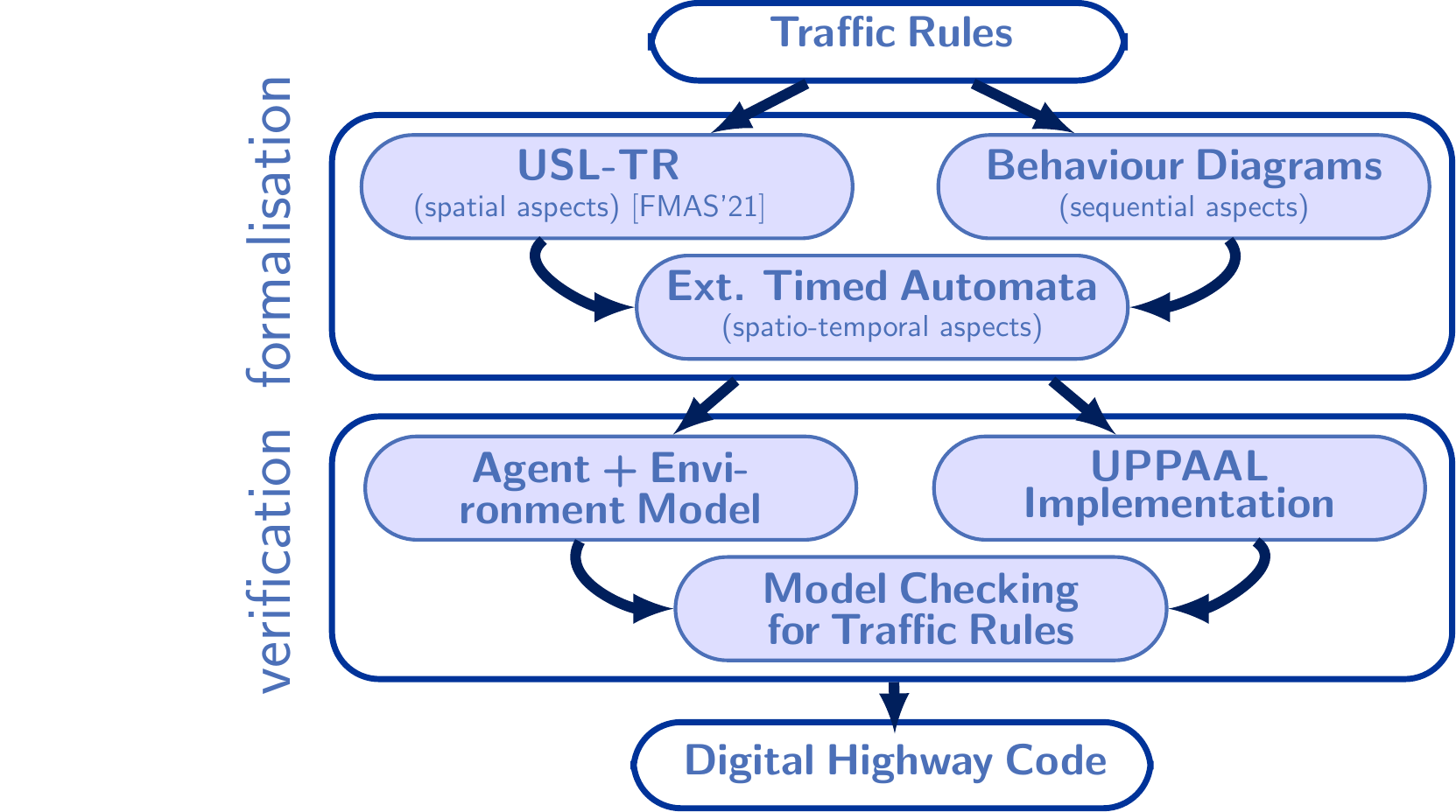}
    \caption{The different steps from natural language traffic rules towards a Digital Highway Code.}
    \label{fig:steps}
\end{figure}

We identify several necessary steps to get from natural language traffic rules to a Digital Highway Code. For this, we start with steps towards a formalisation of traffic rules, followed by some steps towards a verification of the formalised rules, as we visualise in Fig.~\ref{fig:steps}.
For a structural transition of natural language traffic rules to a formalised, digital, and machine-readable, format, we identify two characteristics in each traffic rule: \emph{spatial} and \emph{temporal aspects}. Examples for such spatial elements comprise that there is a traffic sign or a road junction ahead of an AV, or that a safe gap exists to move into. On the other hand, temporal aspects comprise timing behaviour (e.g. that an action does not happen immediately) and the sequentiality of a traffic rule, where, e.g., for changing a lane on a motorway, a car must first check that a sufficiently large free safe gap exists to change into, then set a turn signal, and only then change the lane.

For formalising spatial aspects of traffic rules, in previous work \cite{SchwammbergerAlves21}, we introduce the traffic logic \emph{Urban Spatial Logic for Traffic Rules (USL-TR)}.
In this research preview, we focus on the temporal aspects of traffic rules. We use selected rules from the UK Highway Code \cite{departmentForTransportUsing2017} as examples. Our formalisation of traffic rules follows two steps; first, we translate the traffic rules to behavioural diagrams, i.e. UML activity diagrams \cite{uml}. With this, we capture the sequential aspects of traffic rules (e.g. that an action that is taken after another action or event). Note that these behaviour diagrams do not yet include formal aspects and are an intermediate step between an informal, often ambiguous, natural language traffic rule and a fully formalised traffic rule that is understandable by an agent steering an AV.

Next, we annotate the behaviour diagrams with formulae of USL-TR as guards and invariants to include the spatial traffic aspects into the behaviour diagrams. Further on, we add clock constraints and clock resets to specify timing behaviour of the traffic rule. The result is an extended timed automaton \cite{AD94}, representing both the spatial and the temporal aspects of the respective traffic rule.
With this, we add a formalisation layer to traffic rules, allowing for extracting existent elements from them. Also, this formalised version of traffic rules allows for different verification approaches.

We not only strive to formalise traffic rules, but also to analyse and verify them. With this, we can, for instance, identify conflicting traffic rules and show that autonomous agents can actually safely follow a Digital Highway Code.
For the verification part, we suggest two directions: i) implement the timed automata representations directly into the model-checking tool UPPAAL \cite{BDL04} and ii), enhance the timed automata traffic rules with an agent and environment model.
For the first direction, i), we suggest a parallel composition of different traffic rules to identify potentially conflicting traffic rules. For this direction, we intend to adapt an existing UPPAAL implementation of a timed automaton crossing controller that uses a predecessor logic of USL-TR \cite{BS19}.
For the second direction, ii), synchronisation channels are added to model the communication between the agent and the environment. Thus, we are not only considering the single agent endowed with Traffic Rules but also a necessary adaptation so that this agent works as an AV endowed with a Digital Highway Code, capable of communicating with the corresponding environment. Using this agent and environment model, we envision a formal verification of agents endowed with traffic rules that follows \cite{alvesformalisation2020, jsan10030041} by using the MCAPL framework \cite{Dennis:2012} and the agent programming language \textsc{Gwendolen} \cite{dennis17gwen}. 
We aim to provide a formalised and verifiable Digital Highway Code with these two directions.

Related approaches include the so-called \emph{Traffic Sequence Charts (TSCs)} \cite{DMPR18}. TSCs are based on \emph{Life Sequence Charts (LSCs)} \cite{DH01}, thus also capable of describing dynamic system behaviour. An advantage of our approach, in comparison to TSCs, is that our timed automaton-based traffic rule formalisation allows the use of existing verification techniques like model-checking with UPPAAL. Next, while the industrial standardisation effort \emph{OpenSCENARIO} \cite{OS} allows for the specification of dynamic traffic situations for simulation efforts, it lacks a formal semantics. In contrast, the timed automata specifications that we derive with our approach support the formal analysis of traffic rules.

In Sect.~\ref{sec:pre:usl-tr}, we briefly introduce the logic USL-TR and in Sect.~\ref{sec:behaviour-models}, we exemplarily show respective behaviour diagrams that we build for two of the traffic rules from the UK Highway Code. We add timing behaviour and USL-TR guards to the behaviour diagrams to derive our timed automata in Sect.~\ref{sec:timed-automata}. In Sect.~\ref{sec:vision}, we give an outlook over the steps that are contained within the ``verification'' node of Fig.~\ref{fig:steps}.

\section{Spatial Aspects: USL-TR}\label{sec:pre:usl-tr}
In previous work \cite{SchwammbergerAlves21}, we introduced \emph{Urban Spatial Logic for Traffic Rules (USL-TR)} for reasoning about traffic manoeuvres and rules at urban intersections. USL-TR is a spatial logic and is an extension of \emph{Urban Multi-lane Spatial Logic (UMLSL)} that was introduced in \cite{Sch18-TCS}.

\begin{figure}[htbp]
\centering
    \includegraphics[width=4.8cm, trim= 2.3cm 1.25cm 4.3cm 2cm, clip]{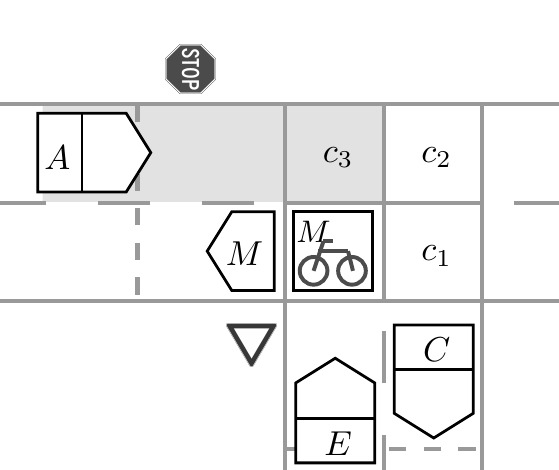}
    \caption{Exemplary traffic situation.}
    \label{fig:traffic-situation}
\end{figure}

Using USL-TR, one can specify spatial features at intersections, e.g. that the space in front of a car $A$ is free in the traffic situation that is depicted in Fig.~\ref{fig:traffic-situation}: $re (A) \frown \mathit{free}$. The atom $re(E)$ comprises the so-called reservation of space for car $E$. Moreover, the operator $\frown$ is inspired by the chop operator from Interval Temporal Logic \cite{Mos85} and splits one larger horizontal interval of space into two smaller ones. Consider the following more complex formula $\mathit{sg}_I (A)$:
\[
    sg_I (A) \,\equiv\, \langle (re(A) \land \neg cs) \,\frown\, (\mathit{free} \land \neg cs) \,\frown\, (\mathit{sg} (A) \land cs) \rangle \text{,}
\]
The formula $\mathit{sg}_I (A)$ is three-parted: we specify that first, there is a reservation of car $A$ not yet on the intersection ($re(E) \land \neg cs$) followed by an interval of free space, also not yet on the intersection ($\mathit{free} \land \neg cs$), followed by a free safe gap on the intersection that is large enough for car $A$ ($\mathit{sg} (A) \land cs$). The formula $\mathit{sg}_I (A)$ holds in the shaded part of Fig.~\ref{fig:traffic-situation}. The abbreviation $\mathit{sg}(A)$ itself comprises the formula $\mathit{sg}(A) \,\equiv\, \mathit{free} \,\land\, \ell >= \mathit{size}_A$, stating that there is a free space greater or equals the size of car $A$. For more details on USL-TR, we refer to \cite{SchwammbergerAlves21}.

Note that with the syntactical element $\ell >= \mathit{size}_A $, we leave the actual size $\mathit{size}_A$ of the safe gap for car $A$ abstract at this point of the specification. The actual semantics of $\mathit{size}_A$ depends on a variety of factors, e.g., the type of the road (asphalt, sand, ...), the weather (sunny, heavy rain, fog, ...) and the power of the brakes. As there exist specific traffic rules for safe gaps for such special cases, it will be of interest to refine the semantics of a safe gap at a later point in our work. An approach to connect a predecessor logic of the spatial logic USL-TR with continuous models and closer-to-reality variable valuations is provided in \cite{ORW17}.\label{page-safe-gap}

\section{Sequential Aspects: Behaviour Diagrams}\label{sec:behaviour-models}
With USL-TR, we can formally specify a variety of spatial aspects of traffic rules. As shown in Fig.~\ref{fig:steps}, our next step is to specify the sequential aspects that are generally contained within a traffic rule. For this, behaviour diagrams, e.g. UML activity diagrams \cite{uml} or simple flowcharts \cite{flowchart}, are a favourable choice, as they allow to model dynamic and sequential system behaviour.
In our case, such dynamic behaviour includes checking for safe gaps, while simultaneously watching out for other road users or reacting to dynamic traffic evolutions. Often, the behaviour that is described within a traffic rule also involves choices to be made, e.g. whether an intersection is entered or not.
In this section, we showcase examples of how to extract behaviour diagrams from the UK Highway Code rules 170 and 171 \cite{departmentForTransportUsing2017}. For these two rules, we previously formalised their spatial aspects using USL-TR in \cite{SchwammbergerAlves21}. In both rules, the overall goal of the AV is to safely enter a road junction.

\textbf{Rule 170.}
We summarise this rule to the following three parts, thereby condensing cyclists, horse riders, motorcyclists and so forth under the general term of ``road users'' (cf.~\cite{departmentForTransportUsing2017}): a) You should watch out for road users. b) Watch out for pedestrians crossing a road junction into which you are turning. If they have started to cross they have priority, so give way. and c) Do not cross or join a road until there is a (safe) gap large enough for you to do so safely.

\begin{figure}[htbp]
    \centering
    \includegraphics[width=0.9\textwidth]{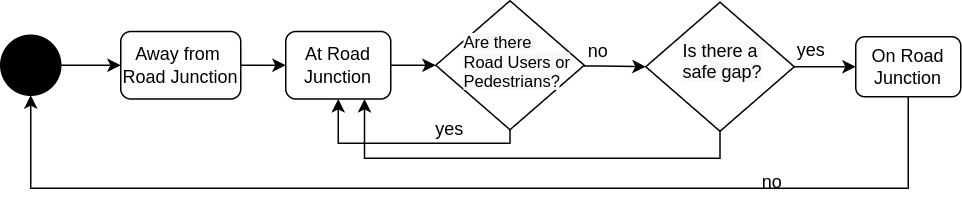}
    \caption{Rule 170: Behaviour Diagram}
    \label{fig:BD-rule170}
\end{figure}

In Fig.~\ref{fig:BD-rule170}, we present the corresponding behaviour diagram for Rule 170, which describes the sequence of actions and choices that are included into rule 170. We identify a sequence of three states:
\begin{itemize}
    \item ``Away from Road Junction'': This is the initial state, where rule 170 does not yet apply, as no road junction is close.
    \item ``At Road Junction'': When the AV arrives at an intersection, some actions are necessary.
    \item ``On Road Junction'': After all parts of rule 170 have been applied, the AV can finally enter the intersection.
\end{itemize}

Note that in our meaning an AV is ``at the road junction'', if it is closer to it than a distance constant $d_c$. This distance $d_c$ depends on different factors, similarly to what we observed for the safe gap earlier on p.~\pageref{page-safe-gap}.
While moving from ``Away from Road Junction'' to ``At Road Junction'' happens immediately (i.e. upon crossing that distance threshold), between the next states some choices need to be made; parts 1 and 2 of rule 170 specify to watch out for both pedestrians and road users (e.g. other AVs, cyclists, ...) and part 3 states that a road junction may only be entered, if there is a safe gap large enough to do so.

\begin{figure}[ht]
    \centering
    \includegraphics[width=\textwidth]{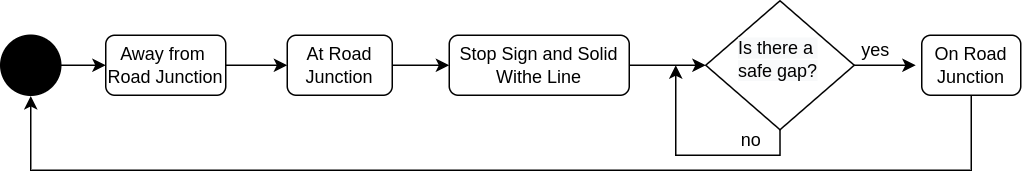}
    \caption{Rule 171: Behaviour Diagram}
    \label{fig:BD-rule171}
\end{figure}

\textbf{Rule 171.}
The rule reads as follows: ``You must stop behind the line at a junction with a `Stop' sign and a solid white line across the road. Wait for a safe gap in the traffic before you move off'' \cite{departmentForTransportUsing2017}.
Fig.~\ref{fig:BD-rule171} shows the behaviour diagram we construct for rule 171. The sequence of main states is the same as for the rule before, with the AV first being away from the road junction and then lastly being on the road junction.
Between the first and the last state, a combined traffic sign needs to be noticed: A stop sign with a solid white line on the road. After that, the AV again needs to check for a safe gap before it may enter the road junction.
We understand that it would be possible (in a further step) to combine the diagrams from Figures~\ref{fig:BD-rule170} and~\ref{fig:BD-rule171} into a single diagram since they share some states. However, we firstly wish to describe each rule separately.

\section{Spatio-Temporal Aspects: Extended Timed Automata}\label{sec:timed-automata}
We have captured the spatial elements of traffic rules using USL-TR (cf.~\cite{SchwammbergerAlves21}) and the sequential aspects using behaviour diagrams (cf.~previous section). We now add time constraints to the behaviour diagrams to model all temporal aspects of traffic rules. We also add the spatial USL-TR aspects to the diagrams so that the resulting model is a timed automaton \cite{AD94} that uses USL-TR formulae as guards and invariants. Thus, this timed automaton captures both spatial and temporal aspects of traffic rules. Such a type of extended timed automaton that uses USL-TR's predecessor logic UMLSL as guards and invariants, as well as particular actions for traffic manoeuvres, has already been introduced in previous work \cite{Sch18-TCS}. This automaton type is named by \emph{Automotive-Controlling Timed Automaton (ACTA)}.

\begin{figure}[htbp]
    \centering
    \includegraphics[width=\textwidth]{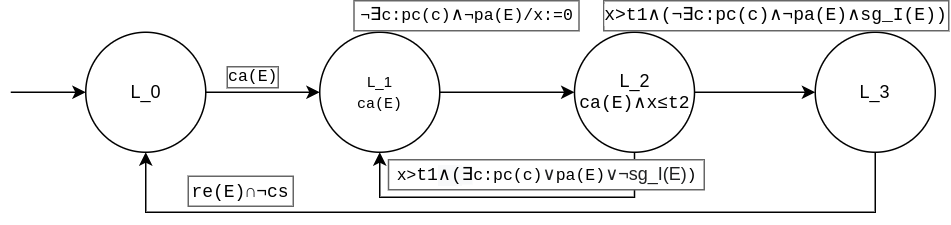}
    \caption{Rule 170: Timed Automaton}
    \label{fig:TA-rule170}
\end{figure}

\textbf{Rule 170.}
In Fig.~\ref{fig:TA-rule170}, we show the timed automaton that corresponds to the previously introduced behaviour diagram (cf.~Fig.~\ref{fig:BD-rule170}).
This timed automaton contains a set of locations ($L_{0}$, $L_{1}$, $\dots$) that mainly conform to the states from the behaviour diagram; 
only the state $L_{2}$ represents a decision process: ``Is there a safe gap?'' (previously seen in Fig.~\ref{fig:BD-rule170}).

In timed automata, taking a transition is a process that happens immediately, without any passing of time. However, in reality, most actions and decisions of an AV will take time, even if it may only be some milliseconds.
Thus, we add a clock variable $x$ to our timed automaton for rule 170 to specify that checking for a safe gap for pedestrians and road users takes some time (from the time interval $[t_1, t_2]$) and does not happen immediately.

The USL-TR guard $\neg \exists c: pc(c) \land \neg pa(E)$ on the transition from location $L_{1}$ to $L_{2}$ checks that there does not exist a potential collision with any road user ($\neg \exists c: pc(c)$) and that no pedestrian is ahead ($pa(E)$). Only if this guard holds it is possible to move from location $L_{1}$ to $L_{2}$.
In the guard on the transition to location $L_{3}$, we add the check for a safe gap on an intersection $sg_I (E)$ to the guards that we already described for the previous transition. This assures that the location $L_{3}$ is only entered if the conditions above are satisfied and additionally there exists a safe gap large enough for the car $E$ on the road junction.
If any of these conditions do not hold, the transition back to location $L_{1}$ is taken, and all the
checks will be done again before the AV may safely enter the road junction.

\begin{figure}[htbp]
    \centering
    \includegraphics[width=0.95\textwidth]{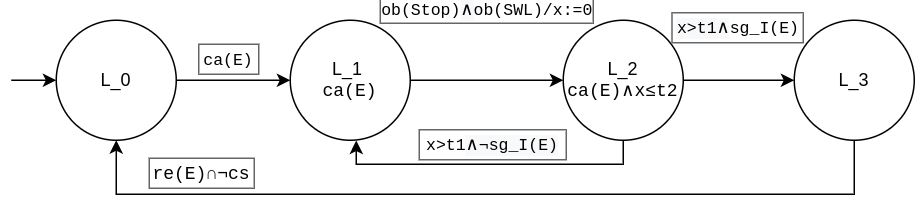}
    \caption{Rule 171: Timed Automaton}
    \label{fig:TA-rule171}
\end{figure}

\textbf{Rule 171.}
Fig.\ref{fig:TA-rule171} depicts the timed automaton that we construct for Rule 171. Notice that to move from location $L_{1}$ to $L_{2}$ the guard $ob(Stop) \land ob(SWL)$ should be satisfied, i.e. the two traffic signs ``Stop'' and ``Solid With Line'' need to be acknowledged by the AV. We again add a clock variable $x$ and clock constraints ($t_1, t_2$) to model that stopping at the traffic signs and checking for a safe gap does not happen immediately. Again, the AV may only move onto the intersection if a safe gap in the intersection has been identified ($sg_I (E)$).

In this section, we introduced extended timed automata representing the UK Highway Code traffic rules 170 and 171. In future work, we intend to formalise further traffic rules from the UK Highway Code, but also driving regulations from other countries (e.g. from the German ``Straßenverkehrsordnung'' \cite{stvo}) in the same manner. For now, we leave the actions of the AVs abstract (e.g. stopping at a traffic sign or setting a turn signal). However, the ACTA formalism from \cite{Sch18-TCS} allows adding such controller actions. Equally, for the agent and environment models that we intend to build for the verification, we will include these actions. 
In the next section, we describe these future steps toward our agent and traffic rule verification endeavour.

\section{On the Verification of Traffic Rules}\label{sec:vision}

Following our diagram from Fig.~\ref{fig:steps}, starting from the timed automata that we obtained in the previous section, we intend to combine two different methods for the verification of traffic rules:
\begin{itemize}
    \item Checking for conflicts and inconsistencies between traffic rules via UPPAAL model-checking \cite{BDL04}
    \item Adding agent and environment models for the verification of traffic-rule-following agents.
\end{itemize}

\textbf{Identifying conflicts and inconsistencies with UPPAAL.}
Traffic rules from the known driving regulations are prone to having inconsistencies and conflicts with each other. Consider the following example: worldwide, most driving regulations contain a rule 
stating that cars must stop at a red traffic light. However, many driving regulations also include the traffic rule of the ``green arrow for turning right'' (``left'' in case of the UK Highway Code). 
This rule allows turning right at a red traffic light if there is enough space. 
Generally, these two rules might be in conflict with one another, as one allows something that the other rule forbids. To altogether find such conflicts, we intend to analyse parallel compositions of traffic rule timed automata (including those that we introduce in Sect.\ref{sec:timed-automata}). 
We will start from a UPPAAL implementation of a protocol for performing turn manoeuvres at intersections from previous work \cite{BS19}, which is modelled as an ACTA and is close to the timed automaton for rule 170 (cf.~Fig.~\ref{fig:TA-rule170}), but it does not yet precisely follow traffic rules.
 After finding conflicting rules with the help of UPPAAL, we need to consider how AVs may solve such conflicts. 
 While humans use common sense for this task,  we envision a situation-dependent prioritisation of AV traffic rules.
 
\textbf{Agent and Environment Models.}
We intend to obtain the corresponding Agent and Environment models from the extended timed automata. 
When getting the Agent model, it is  possible to 
translate it into a sketch of a \emph{Belief-Desire-Intention (BDI)} agent \cite{bratman:87a}.
A BDI agent is defined by a set of plans, where each plan usually comprises the following: 
\textbf{i.} a trigger event;
\textbf{ii.} guards
\textbf{iii.} and a set of actions.
The first two elements of the BDI agent can be obtained from the guards represented in the timed automata using USL-TR.
E.g., the guard 
$\neg \exists c: pc(c) \land \neg pa(E)$ (cf.~Fig.~\ref{fig:TA-rule170}) would generate a trigger event \verb!~!\texttt{potential-collision} 
and a guard \verb!~!\texttt{pedestrian-ahead} for the BDI agent plan.
The third element can be obtained from the locations in the timed automata. E.g., the location $L_{2}$ is responsible for 
checking for a safe gap; as a result, it would generate 
an action for the BDI agent plan named \texttt{checkForSafeGap}. Wrapping it all together,
we have the following BDI agent plan:

\verb!~potential-collision : ~pedestrian-ahead <- checkForSafeGap;!

We could then endow the agent with traffic rules and implement the corresponding agent's
environment. Therefore, extending the agent-based approach for formally verifying the agent's behaviour against traffic rules previously developed in \cite{jsan10030041}.

\textbf{A Digital Highway Code.}
In this research preview, we sketch a methodology (cf. Fig.~\ref{fig:steps}) 
in which we start with natural language traffic rules and, step by step, develop a formalised and verifiable digital version of traffic regulations. By developing techniques for finding and resolving conflicts between traffic rules and introducing verifiable, traffic rule-following, autonomous agents, we pave the way toward a Digital Highway Code. Such a Code will allow AVs to understand, follow and explain traffic rules, which will be necessary for a successful, safe and desirable joint future for humans and AVs.

However, the research is in an early stage and there are several interesting aspects that we will need to keep in mind for a further development of our approach:
\begin{itemize}
    \item Some traffic rules for humans are obsolete for AVs, because they simply do not apply to AVs or because AVs follow them implicitly, ``by design''. For instance, a traffic rule that forbids checking your smartphone while driving, which invariantly holds for an AV.
    \item A mechanism is needed to handle ``broken'' traffic rules. E.g., if a traffic light is defect or a traffic sign is not recognisable due to graffiti or dirt, another traffic rule or plan of action must exist to prevent the AV from stopping forever.
    \item As we sketched before, if some traffic rules are in conflict, the AV needs to be able to prioritise traffic rules to assess which rule must be obeyed and which rule may be temporarily ignored (so called ``trolley problems'' or  ``moral dilemmas'' cf.~\cite{Bonnefon16}). Such a prioritisation of traffic rules cannot be defined globally, but will be situation-dependend.
\end{itemize}

\bibliographystyle{eptcs}
\bibliography{bib}

\begin{thebibliography}{10}
\providecommand{\bibitemdeclare}[2]{}
\providecommand{\surnamestart}{}
\providecommand{\surnameend}{}
\providecommand{\urlprefix}{Available at }
\providecommand{\url}[1]{\texttt{#1}}
\providecommand{\href}[2]{\texttt{#2}}
\providecommand{\urlalt}[2]{\href{#1}{#2}}
\providecommand{\doi}[1]{doi:\urlalt{https://doi.org/#1}{#1}}
\providecommand{\eprint}[1]{arXiv:\urlalt{https://arxiv.org/abs/#1}{#1}}
\providecommand{\bibinfo}[2]{#2}

\bibitemdeclare{article}{AD94}
\bibitem{AD94}
\bibinfo{author}{Rajeev \surnamestart Alur\surnameend} \&
  \bibinfo{author}{David~L. \surnamestart Dill\surnameend}
  (\bibinfo{year}{1994}): \emph{\bibinfo{title}{A Theory of Timed Automata}}.
\newblock {\slshape \bibinfo{journal}{Theoretical Computer Science}}
  \bibinfo{volume}{126}(\bibinfo{number}{2}), pp. \bibinfo{pages}{183--235},
  \doi{10.1016/0304-3975(94)90010-8}.

\bibitemdeclare{inproceedings}{alvesformalisation2020}
\bibitem{alvesformalisation2020}
\bibinfo{author}{Gleifer~Vaz \surnamestart Alves\surnameend},
  \bibinfo{author}{Louise \surnamestart Dennis\surnameend} \&
  \bibinfo{author}{Michael \surnamestart Fisher\surnameend}
  (\bibinfo{year}{2020}): \emph{\bibinfo{title}{Formalisation and
  {Implementation} of {Road} {Junction} {Rules} on an {Autonomous} {Vehicle}
  {Modelled} as an {Agent}}}.
\newblock In: {\slshape \bibinfo{booktitle}{Formal {Methods}. {FM} 2019
  {International} {Workshops}}}, \bibinfo{series}{Lecture {Notes} in {Computer}
  {Science}}, \bibinfo{publisher}{Springer International Publishing},
  \bibinfo{address}{Cham}, pp. \bibinfo{pages}{217--232},
  \doi{10.1007/978-3-030-54994-7\_16}.

\bibitemdeclare{article}{jsan10030041}
\bibitem{jsan10030041}
\bibinfo{author}{Gleifer~Vaz \surnamestart Alves\surnameend},
  \bibinfo{author}{Louise \surnamestart Dennis\surnameend} \&
  \bibinfo{author}{Michael \surnamestart Fisher\surnameend}
  (\bibinfo{year}{2021}): \emph{\bibinfo{title}{A Double-Level Model Checking
  Approach for an Agent-Based Autonomous Vehicle and Road Junction
  Regulations}}.
\newblock {\slshape \bibinfo{journal}{Journal of Sensor and Actuator Networks}}
  \bibinfo{volume}{10}(\bibinfo{number}{3}), \doi{10.3390/jsan10030041}.
\newblock \urlprefix\url{https://www.mdpi.com/2224-2708/10/3/41}.

\bibitemdeclare{inproceedings}{BDL04}
\bibitem{BDL04}
\bibinfo{author}{Gerd \surnamestart Behrmann\surnameend},
  \bibinfo{author}{Alexandre \surnamestart David\surnameend} \&
  \bibinfo{author}{Kim~G. \surnamestart Larsen\surnameend}
  (\bibinfo{year}{2004}): \emph{\bibinfo{title}{A Tutorial on Uppaal}}.
\newblock In: {\slshape \bibinfo{booktitle}{Formal Methods for the Design of
  Real-Time Systems}}, \bibinfo{publisher}{Springer}, pp.
  \bibinfo{pages}{200--236}, \doi{10.1007/978-3-540-30080-9\_7}.

\bibitemdeclare{inproceedings}{BS19}
\bibitem{BS19}
\bibinfo{author}{Christopher \surnamestart Bischopink\surnameend} \&
  \bibinfo{author}{Maike \surnamestart Schwammberger\surnameend}
  (\bibinfo{year}{2019}): \emph{\bibinfo{title}{Verification of Fair
  Controllers for Urban Traffic Manoeuvres at Intersections}}.
\newblock In: {\slshape \bibinfo{booktitle}{Formal Methods. {FM} 2019
  International Workshops - Porto, Portugal, October 7-11, 2019, Revised
  Selected Papers, Part {I}}}, {\slshape \bibinfo{series}{Lecture Notes in
  Computer Science}} \bibinfo{volume}{12232}, \bibinfo{publisher}{Springer},
  pp. \bibinfo{pages}{249--264}, \doi{10.1007/978-3-030-54994-7\_18}.

\bibitemdeclare{article}{Bonnefon16}
\bibitem{Bonnefon16}
\bibinfo{author}{Jean-Fran{\c c}ois \surnamestart Bonnefon\surnameend},
  \bibinfo{author}{Azim \surnamestart Shariff\surnameend} \&
  \bibinfo{author}{Iyad \surnamestart Rahwan\surnameend}
  (\bibinfo{year}{2016}): \emph{\bibinfo{title}{The social dilemma of
  autonomous vehicles}}.
\newblock {\slshape \bibinfo{journal}{Science}}
  \bibinfo{volume}{352}(\bibinfo{number}{6293}), pp.
  \bibinfo{pages}{1573--1576}, \doi{10.1126/science.aaf2654}.
\newblock
  \eprint{http://science.sciencemag.org/content/352/6293/1573.full.pdf}.

\bibitemdeclare{book}{bratman:87a}
\bibitem{bratman:87a}
\bibinfo{author}{Michael~E. \surnamestart Bratman\surnameend}
  (\bibinfo{year}{1987}): \emph{\bibinfo{title}{Intentions, Plans, and
  Practical Reason}}.
\newblock \bibinfo{publisher}{Harvard University Press}, \doi{10.2307/2185304}.

\bibitemdeclare{misc}{stvo}
\bibitem{stvo}
\bibinfo{author}{\surnamestart {Bundesrepublik Deutschland}\surnameend}
  (\bibinfo{year}{2013}): \emph{\bibinfo{title}{Straßenverkehrsordnung
  (StVO)}}.
\newblock
  \urlprefix\url{https://www.gesetze-im-internet.de/stvo_2013/index.html}.

\bibitemdeclare{article}{DH01}
\bibitem{DH01}
\bibinfo{author}{Werner \surnamestart Damm\surnameend} \&
  \bibinfo{author}{David \surnamestart Harel\surnameend}
  (\bibinfo{year}{2001}): \emph{\bibinfo{title}{{LSCs}: Breathing Life into
  Message Sequence Charts}}.
\newblock {\slshape \bibinfo{journal}{Formal Methods in System Design}}
  \bibinfo{volume}{19}(\bibinfo{number}{1}), pp. \bibinfo{pages}{45--80},
  \doi{10.1023/A:1011227529550}.

\bibitemdeclare{inproceedings}{DMPR18}
\bibitem{DMPR18}
\bibinfo{author}{Werner \surnamestart Damm\surnameend}, \bibinfo{author}{Eike
  \surnamestart M{\"{o}}hlmann\surnameend}, \bibinfo{author}{Thomas
  \surnamestart Peikenkamp\surnameend} \& \bibinfo{author}{Astrid \surnamestart
  Rakow\surnameend} (\bibinfo{year}{2018}): \emph{\bibinfo{title}{A Formal
  Semantics for Traffic Sequence Charts}}.
\newblock In: {\slshape \bibinfo{booktitle}{Principles of Modeling - Essays
  Dedicated to Edward A. Lee on the Occasion of His 60th Birthday}},
  \bibinfo{publisher}{Springer}, pp. \bibinfo{pages}{182--205},
  \doi{10.1007/978-3-319-95246-8\_11}.

\bibitemdeclare{techreport}{dennis17gwen}
\bibitem{dennis17gwen}
\bibinfo{author}{Louise~A. \surnamestart Dennis\surnameend}
  (\bibinfo{year}{2017}): \emph{\bibinfo{title}{Gwendolen Semantics: 2017}}.
\newblock \bibinfo{type}{Technical Report} \bibinfo{number}{ULCS-17-001},
  \bibinfo{institution}{University of Liverpool, Department of Computer
  Science}.

\bibitemdeclare{article}{Dennis:2012}
\bibitem{Dennis:2012}
\bibinfo{author}{Louise~A. \surnamestart Dennis\surnameend},
  \bibinfo{author}{Michael \surnamestart Fisher\surnameend},
  \bibinfo{author}{Matthew~P. \surnamestart Webster\surnameend} \&
  \bibinfo{author}{Rafael~H. \surnamestart Bordini\surnameend}
  (\bibinfo{year}{2012}): \emph{\bibinfo{title}{{Model Checking Agent
  Programming Languages}}}.
\newblock {\slshape \bibinfo{journal}{Automated Software Engineering}}
  \bibinfo{volume}{19}(\bibinfo{number}{1}), pp. \bibinfo{pages}{5--63},
  \doi{10.1007/s10515-011-0088-x}.

\bibitemdeclare{misc}{departmentForTransportUsing2017}
\bibitem{departmentForTransportUsing2017}
\bibinfo{author}{\surnamestart {Department for Transport}\surnameend}
  (\bibinfo{year}{2017}): \emph{\bibinfo{title}{Using the road (159 to 203) -
  {The} {Highway} {Code} - {Guidance} - {GOV}.{UK}}}.
\newblock
  \urlprefix\url{https://www.gov.uk/guidance/the-highway-code/using-the-road-159-to-203}.

\bibitemdeclare{book}{uml}
\bibitem{uml}
\bibinfo{author}{Martin \surnamestart Fowler\surnameend}
  (\bibinfo{year}{2003}): \emph{\bibinfo{title}{UML Distilled: A Brief Guide to
  the Standard Object Modeling Language}}, \bibinfo{edition}{3} edition.
\newblock \bibinfo{publisher}{Addison-Wesley Longman Publishing Co., Inc.},
  \bibinfo{address}{USA}.

\bibitemdeclare{book}{flowchart}
\bibitem{flowchart}
\bibinfo{author}{Frank~B. \surnamestart Gilbreth\surnameend} \&
  \bibinfo{author}{Lillian~M. \surnamestart Gilbreth\surnameend}
  (\bibinfo{year}{1921}): \emph{\bibinfo{title}{Process Charts}}.
\newblock \bibinfo{publisher}{The American Society of Mechanical Engineers},
  \bibinfo{address}{New York, USA}.
\newblock
  \urlprefix\url{https://www.thegilbreths.com/resources/Gilbreth-Process-Charts-1921.pdf}.

\bibitemdeclare{misc}{OS}
\bibitem{OS}
\bibinfo{author}{IRES~Simulationstechnologie \surnamestart GmbH.\surnameend}
  (\bibinfo{year}{2022}): \emph{\bibinfo{title}{OpenSCENARIO}}.
\newblock \urlprefix\url{http://www.openscenario.org}.
\newblock \bibinfo{note}{Accessed: 2022-08-05}.

\bibitemdeclare{article}{Mos85}
\bibitem{Mos85}
\bibinfo{author}{Ben \surnamestart Moszkowski\surnameend}
  (\bibinfo{year}{1985}): \emph{\bibinfo{title}{A Temporal Logic for Multilevel
  Reasoning About Hardware}}.
\newblock {\slshape \bibinfo{journal}{Computer}}
  \bibinfo{volume}{18}(\bibinfo{number}{2}), pp. \bibinfo{pages}{10--19},
  \doi{10.1109/MC.1985.1662795}.

\bibitemdeclare{incollection}{ORW17}
\bibitem{ORW17}
\bibinfo{author}{Ernst{-}R{\"{u}}diger \surnamestart Olderog\surnameend},
  \bibinfo{author}{Anders~P. \surnamestart Ravn\surnameend} \&
  \bibinfo{author}{Rafael \surnamestart Wisniewski\surnameend}
  (\bibinfo{year}{2017}): \emph{\bibinfo{title}{Linking Discrete and Continuous
  Models, Applied to Traffic Manoeuvrers}}.
\newblock In \bibinfo{editor}{Michael~G. \surnamestart Hinchey\surnameend},
  \bibinfo{editor}{Jonathan~P. \surnamestart Bowen\surnameend} \&
  \bibinfo{editor}{Ernst{-}R{\"{u}}diger \surnamestart Olderog\surnameend},
  editors: {\slshape \bibinfo{booktitle}{Provably Correct Systems}},
  \bibinfo{series}{{NASA} Monographs in Systems and Software Engineering},
  \bibinfo{publisher}{Springer}, pp. \bibinfo{pages}{95--120},
  \doi{10.1007/978-3-319-48628-4\_5}.

\bibitemdeclare{article}{Sch18-TCS}
\bibitem{Sch18-TCS}
\bibinfo{author}{Maike \surnamestart Schwammberger\surnameend}
  (\bibinfo{year}{2018}): \emph{\bibinfo{title}{An abstract model for proving
  safety of autonomous urban traffic}}.
\newblock {\slshape \bibinfo{journal}{Theoretical Computer Science}}
  \bibinfo{volume}{744}, pp. \bibinfo{pages}{143--169},
  \doi{10.1016/j.tcs.2018.05.028}.

\bibitemdeclare{inproceedings}{SchwammbergerAlves21}
\bibitem{SchwammbergerAlves21}
\bibinfo{author}{Maike \surnamestart Schwammberger\surnameend} \&
  \bibinfo{author}{Gleifer~Vaz \surnamestart Alves\surnameend}
  (\bibinfo{year}{2021}): \emph{\bibinfo{title}{Extending Urban Multi-Lane
  Spatial Logic to Formalise Road Junction Rules}}.
\newblock In: {\slshape \bibinfo{booktitle}{Proceedings Third Workshop on
  Formal Methods for Autonomous Systems, {FMAS} 2021, Virtual, 21st-22nd of
  October 2021}}, {\slshape \bibinfo{series}{{EPTCS}}} \bibinfo{volume}{348},
  pp. \bibinfo{pages}{1--19}, \doi{10.4204/EPTCS.348.1}.

\end{thebibliography}

\end{document}